\pdfoutput=1

\documentclass[12pt,a4paper,hyperref]{JHEP3}
\let\ifpdf\relax
\usepackage{ifpdf}
\usepackage{color}
\let\normalcolor\relax
\usepackage{mathtools}
\usepackage{cite}
\newcommand\sss{\hspace{0.7pt}}
\newcommand{\stwo}[2]{s_{#1\sss #2}}
\newcommand{\sthree}[3]{s_{#1\sss #2 \sss #3}}
\newcommand{\sfour}[4]{s_{#1\sss #2 \sss #3\sss #4}}
\newcommand{\sfive}[5]{s_{#1\sss #2 \sss #3\sss #4 \sss #5}}
\newcommand{\ssix}[6]{s_{#1\sss #2 \sss #3\sss #4 \sss #5\sss #6}}
\renewcommand{\phi}{\varphi}
\newcommand{\chyInt}{d\Omega_{\mathrm{CHY}}}
\newcommand{\mi}{\raisebox{0.75pt}{\scalebox{0.75}{$\,-\,$}}}
\newcommand{\pl}{\raisebox{0.75pt}{\scalebox{0.75}{$\,+\,$}}}
\newcommand{\LR}{\raisebox{-2.25pt}{\,\scalebox{1.75}{$\Leftrightarrow$}}\,}
\newcommand{\Rarrow}{\raisebox{-2.25pt}{\,\scalebox{1.75}{$\Rightarrow$}}\,}
\newcommand{\bbox}[1]{\raisebox{-2.25pt}{\,\scalebox{1.75}{$#1$}}\,}
\newcommand{\fwbox}[2]{\text{\makebox[#1][c]{$\hspace{-150pt}\displaystyle#2\hspace{-150pt}$}}}

\newcommand{\eq}[1]{\vspace{-2.5pt}\begin{equation}\hspace{-100pt}#1\hspace{-100pt}\vspace{-2.5pt}\end{equation}}

\newcommand{\fig}[3]{\raisebox{#1}{\includegraphics[scale=#2]{#3}}}
\newcommand{\Binv}[1]{\mathcal{B}^{}[#1]}
\newcommand{\z}[2]{(z_{#1}\!-\!z_{#2})}
\newcommand{\Pfprime}{\text{Pf}\,'\!\hspace{1pt}}

\preprint{2015}
\title{\mbox{{\LARGE Scattering Equations and Feynman Diagrams}}}
\author{Christian Baadsgaard, N.~E.~J.~Bjerrum-Bohr, Jacob~L.~Bourjaily, and Poul~H.~Damgaard\\
\mbox{{Niels Bohr International Academy and Discovery Center}}\\
\mbox{{Niels Bohr Institute, University of Copenhagen,}}\\
\mbox{Blegdamsvej 17, DK-2100 Copenhagen \O, Denmark}\vspace{-8pt}}
\abstract{
We show a direct matching between individual Feynman diagrams and integration measures in the scattering equation formalism of Cachazo, He and Yuan. The connection is most easily explained in terms of triangular graphs associated with planar Feynman diagrams in $\phi^3$-theory. We also discuss the generalization to general scalar field theories with $\phi^p$ interactions, corresponding to polygonal graphs involving vertices of order $p$. Finally, we describe how the same graph-theoretic language can be used to provide the precise link between individual Feynman diagrams and string theory integrands.
}

\begin{document}\newpage
\section{Introduction}\label{sec:introduction}
It was recently shown by Cachazo, He and Yuan (CHY) that for many theories, tree-level scattering 
amplitudes involving any number of massless fields in any number of dimensions can be represented in terms of auxiliary variables $z_i$ constrained by the so-called scattering equations, \cite{Cachazo:2013gna,Cachazo:2013hca,Cachazo:2013iea,Cachazo:2014xea}.
For $n$-point kinematics, the scattering equations can be written,
\vspace{10pt}\eq{\sum_{i\neq j} \frac{\stwo{i}{j}}{\z{i}{j}}\;=\;0\,,\label{scattering_equations}} 
where $\stwo{i}{j}\!=\!(p_i+p_j)^2$ are ordinary Mandelstam variables. In general, there are $(n\mi3)!$ solutions to the scattering equations (found after eliminating the $SL(2,\mathbb{C})$ redundancy in the $z_i$ variables). For a fairly broad class of theories, scattering amplitudes can be represented in the CHY formalism as integrals over the $z_i$ variables, fully localized by the constraints, (\ref{scattering_equations}). A proof of the CHY construction of $\phi^3$-theory and Yang-Mills theory was given by Dolan and Goddard in \mbox{ref.\ \cite{Dolan:2013isa}}. 

A close relationship between the CHY representations and string theory was early noted. Indeed, a reinterpretation of
the CHY prescription in terms of a complexified worldline (in the infinite tension limit)
was described by Mason and Skinner in \mbox{ref.\ \cite{Mason:2013sva}} and by 
Berkovitz in \mbox{ref.\ \cite{Berkovits:2013xba}} (see also \mbox{refs.\ \cite{Gomez:2013wza,
Adamo:2013tsa, Ohmori:2015sha}}). In a similar manner, conventional superstring theory can be linked 
directly to the CHY formalism after manifest cancellation of tachyon poles in the superstring
integrand as described in \mbox{ref.\ \cite{Bjerrum-Bohr:2014qwa}}, which extended the use of the scattering equation formalism to a number of theories beyond those considered by Cachazo, He and Yuan---for example, scattering amplitudes involving mixed particle types including fermions. The precise link between the two types of integrands follows as a corollary of a more general set of integration rules recently described in \mbox{ref.\ \cite{Baadsgaard:2015voa}}.

In this paper, we wish to take the analysis of \mbox{ref.\ \cite{Baadsgaard:2015voa}} one step further: to use the rules for integration described in \cite{Baadsgaard:2015voa} to systematically construct integrands for individual Feynman diagrams. As expected, this is most easily
done in the case of $\phi^3$-theory, where we can make use of a simple graph-theoretic map
between any  Feynman diagram, and a triangular covering of the graph. This graph of triangles
yields a direct translation into the correct CHY integrand. And this construction naturally generalizes: for Feynman
diagrams involving any combination of (possibly mixed) higher-order vertices, there exists a corresponding polygonal graph from which one can read-off the corresponding CHY integrand. The
only complication for theories involving higher than cubic vertices is that the representation will require normalization factors that depend on the external momenta.
We provide a systematic prescription for how to determine these normalization factors below.

Our paper is organized as follows. We review the scattering equations and how to use their solutions to provide representations of tree-level scattering amplitudes in \mbox{section \ref{sec:newrules}}, briefly summarizing the integration rules described in \mbox{ref.\ \cite{Baadsgaard:2015voa}}. In \mbox{section \ref{sec:weavingdiagrams}}, we recast the original scattering equation formalism for $\phi^3$-theory in a diagrammatic manner which allows us to relate individual Feynman diagrams (and sums of Feynman diagrams) directly to CHY integrands. By invoking the integration rules of \mbox{ref. \cite{Baadsgaard:2015voa}}, we are able to prove the polygon decomposition first put forward in \cite{Cachazo:2013iea}. And we discuss how this generalizes to represent scattering amplitudes in $\phi^p$-theory (including theories with mixed orders) in \mbox{section \ref{sec:phi_p_theories}}. Finally, we comment on the corresponding analysis in string theory together with our conclusions in \mbox{section \ref{sec:conclusions}}.

\section{\mbox{Scattering Equation Constraints, and Rules for Integration}}\label{sec:newrules}
Let us briefly review the new diagrammatic rules for computing integrals in the scattering equation formalism. 
In the scattering equation framework, the $n$-point scattering amplitude in $\phi^3$-theory may be represented,
\eq{\mathcal{A}_n^{\phi^3}=\int\!\chyInt\,\,
\left(\frac{1}{\z{1}{2}^2\z{2}{3}^2\!\cdots\z{n}{1}^2}\right)\,,\label{n_point_function_in_phi3_thy}}
where $\chyInt$ denotes the following integration measure combined with the 
scattering equation constraints,\\
\eq{\!\!\!\!\!\!\!\!\!\!\!\!\!\!\!\chyInt\equiv\frac{d^nz}{\mathrm{vol}(SL(2,\mathbb{C})\!)}
\prod_i\,\!'\delta(S_i)=\!\z{r}{s}^2\z{s}{t}^2\z{t}{r}^2
\prod_{\fwbox{25pt}{i\!\in\!\mathbb{Z}_n\!\backslash\{r,s,t\}}} dz_i\,\delta(S_i)\,,\label{definition_of_chy_measure}}
(independent of the choice of $\{r,s,t\}$), where $S_i$ denotes the $i^{\mathrm{th}}$ scattering equation,
\eq{S_i\equiv\sum_{j\neq i}\frac{\stwo{i}{j}}{\z{i}{j}}\,,}
where $p_i$ and $p_j$ are on-shell so that $\stwo{i}{j}\!\equiv\!(p_i+p_j)^2\!=\!2(p_i\cdot p_j)$; in general, we define
$s_{i \:\! j \:\! \cdots \:\! k}\!\equiv\!(p_i\pl p_j\pl\cdots\pl p_k)^2$. 

Although the scattering equation $\delta$-functions in (\ref{definition_of_chy_measure}) completely localize 
the integral over $\chyInt$, the number of solutions to the scattering equations, $(n\mi3)!$, grows rapidly 
with the number of particles, making all such integrals computationally quite challenging. Conveniently, 
for a very large class of integrands $\mathcal{I}(z)$, there exists a simple, combinatorial rule for determining 
$\int\!\chyInt\,\,\mathcal{I}(z)$ as described in \mbox{ref.\ \cite{Baadsgaard:2015voa}}. We review this rule presently.

Let $\mathcal{I}(z)$ be any integrand involving arbitrary products of factors $\z{i}{j}$ in the denominator, 
subject to the requirement, imposed by M\"obius invariance, that each number 1 to $n$ appears in 
exactly four factors. We can represent the integrand $\mathcal{I}(z)$ diagrammatically as a so-called $4$-regular graph by representing the 
$z_i$'s as the vertices of an $n$-gon, and drawing a single edge between vertices $z_i$ and $z_j$ for 
each factor of $\z{i}{j}$ (with multiplicity) in the denominator of $\mathcal{I}(z)$. (For the sake of concreteness, we will always consider 
the factors $\z{i}{j}$ to be ordered so that $i\!<\!j$ with respect to the (arbitrary) ordering of the labels of the $z_i$'s.) 

The integral $\int\!\chyInt\,\,\mathcal{I}(z)$ will consist of a sum of inverse-products of Mandelstam variables of the form,
\vspace{-2.5pt}\eq{\prod_{a=1}^{n-3}\Big(1/{s_{q_a}}\Big)\,,\vspace{-2.5pt}}
where each subset $q_a\!\subset\!\{1,\ldots,n\}$ has at most $n/2$ elements 
(with $q_a\!\simeq\!q_a^c\!\equiv\!\mathbb{Z}_n\!\backslash q_a$), and for which the collection of subsets $\{q_a\}$ 
satisfy the following criteria: \\[-18pt]
\begin{itemize}
\item for each $q_a$ there are exactly $(2|q_a|\mi2)$ factors $\z{i}{j}$ (including multiplicity) in the denominator 
of $\mathcal{I}(z)$ involving pairs $\{i,j\}\!\subset\! q_a$; \\[-24pt]
\item every pair $\{q_a,q_b\}$ is either nested or complementary---that is, $q_a\!\subset\!q_b$ or $q_b\!\subset\!q_a$, or $q_a\!\subset\!q_b^c$ or $q_b^c\!\subset\!q_a$.\\[-18pt]
\end{itemize} 
If there are no collections of subsets $\{q_a\}$ which satisfy the criteria above, the result of integration will be zero.\footnote{The cases where these rules do not work are those where there is a subset $q_a$ with more than $(2|q_a|\mi2)$ factors $\z{i}{j}$ in the denominator 
of $\mathcal{I}(z)$ involving pairs $\{i,j\}\!\subset\! q_a$. But these cases are not relevant to this paper.}

Although this rule may appear somewhat involved, it is entirely combinatorial and therefore provides a 
simple way to determine the result of integrating an integrand $\mathcal{I}(z)$ against the measure $\chyInt$, which imposes the scattering equations as constraints (including those not immediately obvious from the examples discussed in \mbox{ref.\ \cite{Cachazo:2015nwa}}). 

We can illustrate both the rules described above and the diagrammatic representation of integrands with the following example:
\vspace{-5pt}\eq{\hspace{-20pt}\fig{-36.375pt}{1}{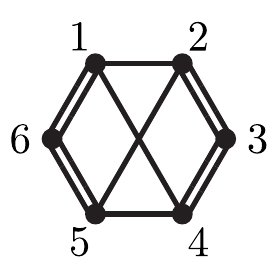}\Rarrow\Bigg(\frac{1}{\stwo{2}{3}\stwo{5}{6}\sthree{5}{6}{1}}
+\frac{1}{\stwo{2}{3}\stwo{6}{1}\sthree{5}{6}{1}}+\frac{1}{\stwo{3}{4}\stwo{5}{6}\sthree{5}{6}{1}}
+\frac{1}{\stwo{3}{4}\stwo{6}{1}\sthree{5}{6}{1}}\Bigg)\,.\vspace{-5pt}\label{6pt_2_1_integration_example}}
Less trivial examples include, 
\vspace{-2.5pt}\eq{\hspace{-30pt}\fig{-36.375pt}{1}{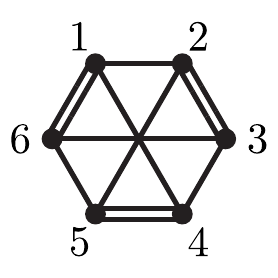}\!\Rarrow\frac{1}{\stwo{2}{3}\stwo{4}{5}\stwo{6}{1}},\hspace{-10pt}\text{}\quad\fig{-36.375pt}{1}{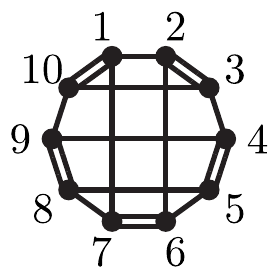}\!\Rarrow\frac{1}{\stwo{2}{3}\stwo{4}{5}\stwo{6}{7}\stwo{8}{9}\stwo{10}{ 1}\sfour{10}{ 1}{2}{3}\sfour{4}{5}{8}{9}}\,.\vspace{-5pt}\label{further_integration_examples}}

It is a relatively simple exercise to see that this rule correctly reproduces the CHY representation of amplitudes in scalar $\phi^3$-theory, (\ref{n_point_function_in_phi3_thy}); many further examples were described in \mbox{ref.\ \cite{Baadsgaard:2015voa}}. In this work, we will mostly be concerned with integrands that correspond to (contributions to) scattering 
amplitudes in scalar field theories.

\section{Feynman Diagrams, Polygon Graphs and CHY Integrands}\label{sec:weavingdiagrams}
It is a proposition in \mbox{ref.\ \cite{Cachazo:2013iea}} that certain specific CHY diagrams can be evaluated by 
decomposing them into polygons. By applying the integration rules of the previous section, it is possible to prove this proposition and provide a direct correspondence between individual $\phi^3$ Feynman diagrams (which may contain $p$-point sub-amplitudes) and specific CHY integrands.

Given a tree-level Feynman diagram in $\phi^3$ scalar field theory, possibly involving $p$-point sub-amplitudes, the corresponding CHY integrand can be constructed as follows. Without loss of generality, we may assume the Feynman diagram is planar (with respect to some ordering of the external legs) and connected. First, replace each $p$-point vertex in the diagram with a $p$-gon whose corners lie along the legs involved (with the polygons of internally-connected vertices meeting at their corners). The resulting, `polygon graph' encodes a single, Hamiltonian cycle that crosses itself wherever two polygons meet, and visits every external leg once. This cycle, together with an $n$-cycle connecting the outer legs of the graph (according to the planar embedding) results in a $4$-regular graph with two Hamiltonian cycles that encodes a CHY integrand. 

This rule can be illustrated as follows:
\vspace{-2.5pt}\eq{\hspace{-10pt}\fwbox{0pt}{\fig{-56.375pt}{1}{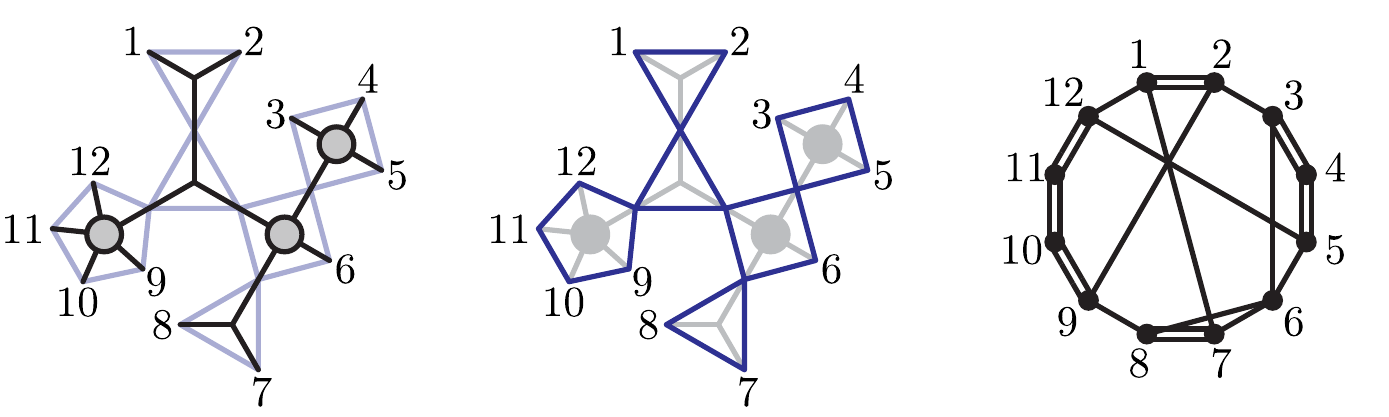}}\fwbox{0pt}{\Rarrow\hspace{120pt}\Rarrow}\vspace{-2.5pt}\label{first_example_polygon_graph}}
where the CHY integrand is represented according to the same conventions of \mbox{ref.\ \cite{Baadsgaard:2015voa}}. It is worth mentioning that there is another way to associate planar Feynman diagrams with $4$-regular graphs: rather than representing each $p$-point vertex in the diagram with a $p$-gon, we could instead represent each with a ``weave'' as in,
\vspace{-5.5pt}\eq{\hspace{20pt}\fwbox{0pt}{\fig{-56.375pt}{1}{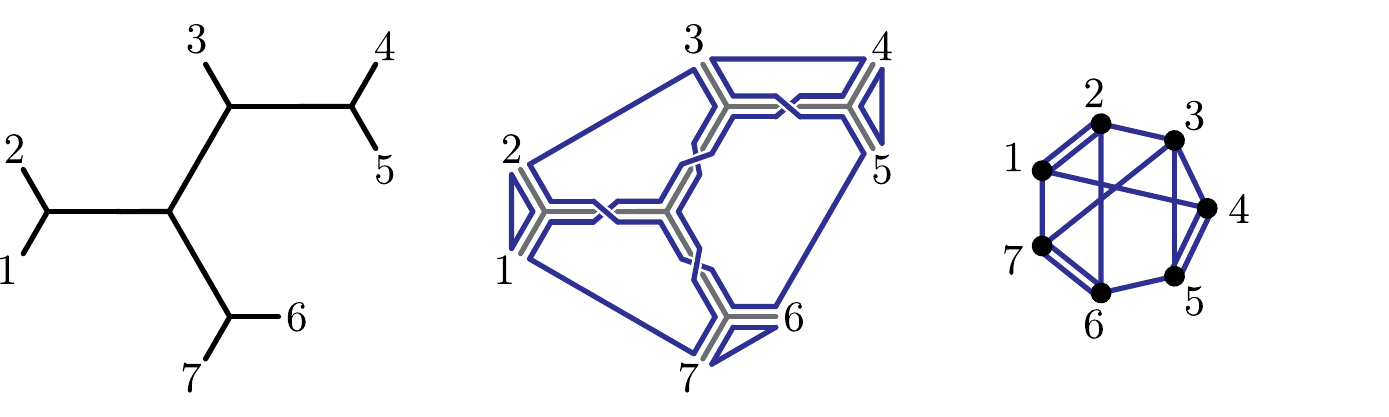}}\fwbox{0pt}{\Rarrow\hspace{120pt}\Rarrow}\vspace{-5pt}\label{second_example_polygon_graph}}%
Such ``weaved'' Feynman diagrams are certainly quite suggestive, but we will mostly use polygon diagrams such as (\ref{first_example_polygon_graph}) below. 

In both examples above, the pair of Hamiltonian cycles---the one following from the $p$-gon vertices, and the one encircling the boundary---overlap in several places, resulting in double-lines in the diagram which encodes the CHY integrand. For the second example, the final $4$-regular graph corresponds to the integrand:
\vspace{-6pt}\eq{\hspace{-26pt}\fig{-36.375pt}{1}{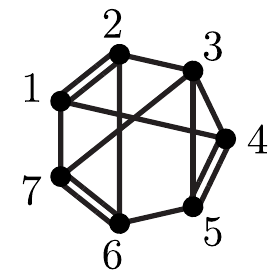}\!\!\LR\!\!\left\{\!\!\begin{array}{@{}l@{}l}&\displaystyle\frac{1}{\z{1}{2}\z{2}{3}\z{3}{4}\z{4}{5}\z{5}{6}\z{6}{7}\z{7}{1}}\\\displaystyle\times&\displaystyle\frac{1}{\z{1}{2}\z{2}{6}\z{6}{7}\z{7}{3}\z{3}{5}\z{5}{4}\z{4}{1}}\,.\end{array}\right.\!\!\!\;\;\label{second_example_integrand}\vspace{-5pt}}
Using the rules of integration described in \mbox{ref.\ \cite{Baadsgaard:2015voa}} and reviewed above, the result of integrating (\ref{second_example_integrand}) with the CHY measure $\chyInt$ would result in,
\vspace{-6pt}\eq{\fig{-36.375pt}{1}{7pt_1_1}\Rarrow\frac{1}{\stwo12\stwo45\stwo67\sthree345}\,.\vspace{-10pt}}

Let us consider another example of a diagram involving $p$-point vertices ($p$-point sub-amplitudes in $\phi^3$-theory). The following graph,
\vspace{-5pt}\eq{\hspace{-30pt}\fwbox{0pt}{\fig{-46.5pt}{1}{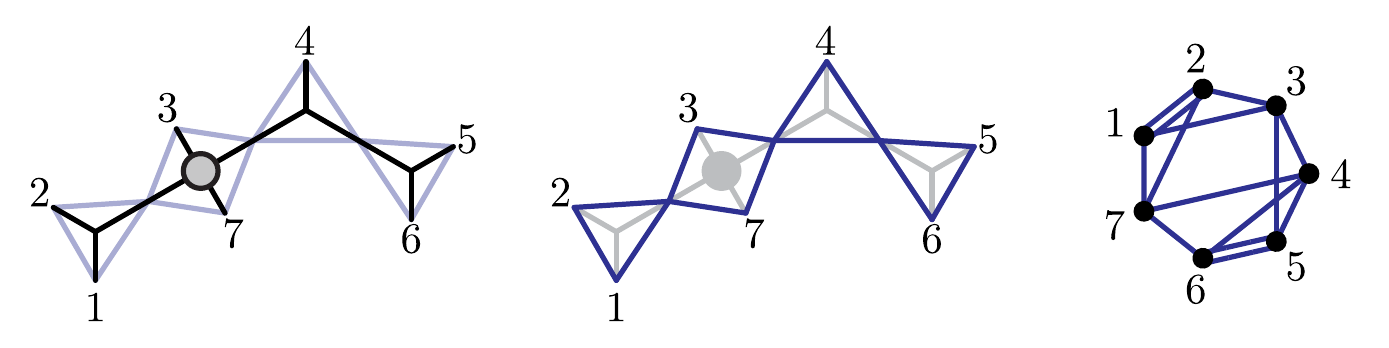}}\fwbox{0pt}{\hspace{45pt}\Rarrow\hspace{130pt}\Rarrow}\vspace{-5pt}}%
would correspond to the CHY integrand,
\vspace{-5pt}\eq{\hspace{-26pt}\fig{-36.375pt}{1}{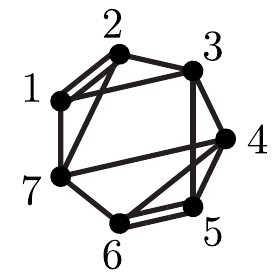}\!\!\LR\!\!\left\{\!\!\begin{array}{@{}l@{}l}&\displaystyle\frac{1}{\z{1}{2}\z{2}{3}\z{3}{4}\z{4}{5}\z{5}{6}\z{6}{7}\z{7}{1}}\\\displaystyle\times&\displaystyle\frac{1}{\z{1}{2}\z{2}{7}\z{7}{4}\z{4}{6}\z{6}{5}\z{5}{3}\z{3}{1}}\,.\end{array}\right.\!\!\!\;\;\vspace{-5pt}}
Because the $4$-point vertex represents the $4$-point amplitude in $\phi^3$-theory, we can replace it with a sum of Feynman diagrams, resulting in the integrand-level identity:
\vspace{-10pt}\eq{\hspace{-24pt}\begin{array}{@{}c@{}}\hspace{-130pt}\fig{-36.375pt}{1}{7pt_2_1}\bbox{=}\fig{-36.375pt}{1}{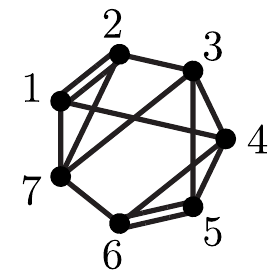}\bbox{+}\!\fig{-36.375pt}{1}{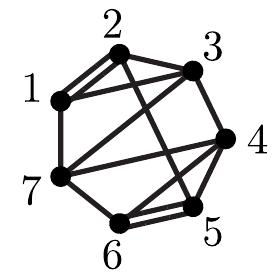}\\\LR\!\!\left\{\begin{array}{@{}l@{}l@{}}
&\displaystyle\frac{1}{\z{1}{2}^2\z{2}{3}\z{3}{4}\z{4}{5}\z{5}{6}^2\z{6}{7}\z{7}{1}}\\[10pt]\displaystyle\times&\displaystyle\frac{1}{\z{3}{7}\z{4}{6}}\left[\frac{1}{\z{1}{4}\z{2}{7}\z{3}{5}}+\frac{1}{\z{2}{5}\z{7}{4}\z{1}{3}}\right]\,.\end{array}\right.\!\!\end{array}\vspace{-15pt}}
\\

\subsection{Proof of the Correspondence with Feynman Graphs}
Let us now demonstrate that the correspondence described above provides a correct representation for all tree-level Feynman diagrams involving arbitrary-order vertices corresponding to $p$-point sub-amplitudes in scalar $\phi^3$-theory. 

First, notice that the rule above reproduces the CHY representation for any $n$-point amplitude in $\phi^3$-theory. In this case, the two Hamiltonian cycles are clearly identical, resulting in the integrand appearing in equation (\ref{n_point_function_in_phi3_thy}). From this, the general claim follows inductively from the consideration of how the rule works when any two diagrams are merged according to:
\eq{\fig{-36.375pt}{1}{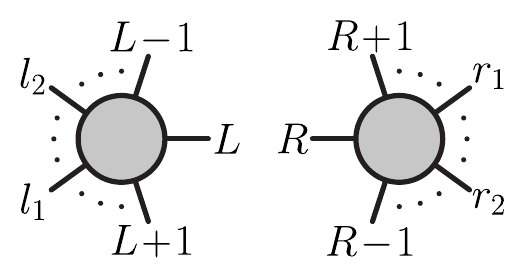}\bbox{\Rightarrow}\fig{-36.375pt}{1}{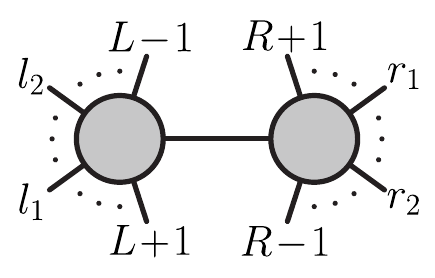}\label{merger_induction_blob_figure}}
According to the rule describe above, merging two graphs corresponds to the following operation on their CHY representations. By induction, each of the subgraphs correspond to CHY integrands associated with $4$-regular diagrams; it is not hard to see that the action of merging two graphs according to (\ref{merger_induction_blob_figure}) results in the following action on the corresponding CHY integrands:
\vspace{-5pt}\eq{\fig{-36.375pt}{1}{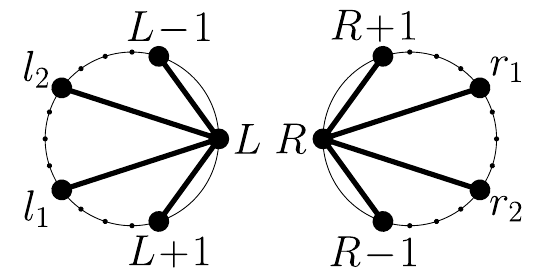}\bbox{\Rightarrow}\fig{-56.375pt}{1}{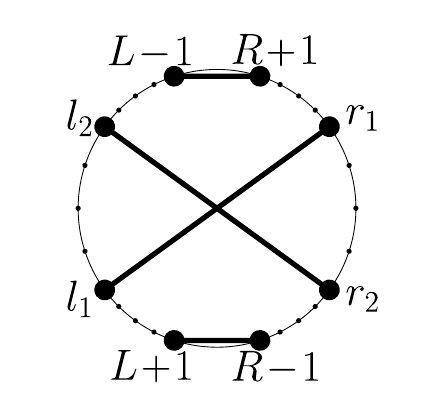}\label{merger_induction_figure}\vspace{-5pt}}

It is a relatively simple exercise to show that the action above correctly preserves the factorization channels of each of the graphs being merged, and introduces exactly one new Mandelstam invariant corresponding to the factorization channel. Moreover, the operation (\ref{merger_induction_figure}) preserves the polygon structure of the constituent CHY diagrams. This demonstrates that the polygon rule described above correctly represents any graph as a CHY integrand. 

Notice that we can use this rule to repeatedly connect polygon vertices of CHY diagrams for $\phi^3$-theory (each encoding a collection of diagrams corresponding to the sub-amplitude), to represent any tree-level Feynman graph. For example, we may sew together $5$-point sub-amplitudes into graphs such as:
\eq{\hspace{-20pt}\begin{array}{@{}c@{}c@{}c@{}}\fwbox{120pt}{\fig{-36.375pt}{1}{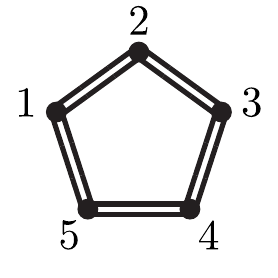}}&\fwbox{140pt}{\fig{-36.375pt}{1}{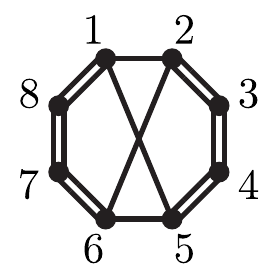}}&\fig{-56.375pt}{1}{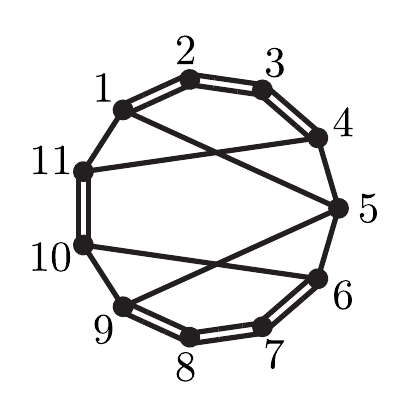}\\\fig{-36.375pt}{1}{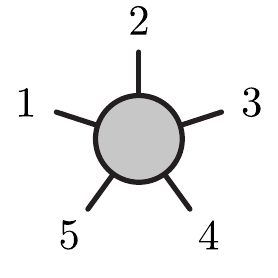}&\fig{-36.375pt}{1}{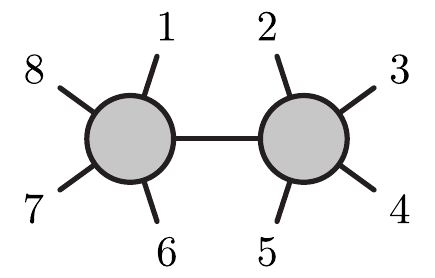}&\fig{-36.375pt}{1}{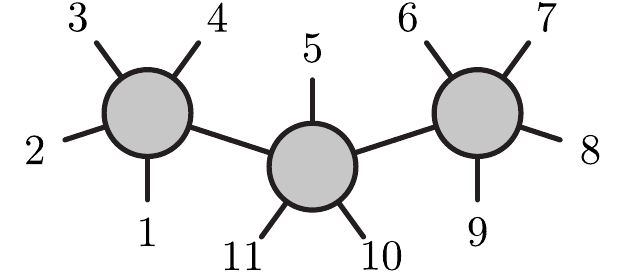}\end{array}\label{phififth}}
We note that this is very close to defining $(3V\pl2)$-point amplitudes in a fundamental $\phi^5$ theory. What is missing is the sum over factorized channels and the correct normalization factors which effectively replace the 5-point sub-amplitudes in $\phi^3$-theory with fundamental vertices. We will return to this in detail in the next section.

\subsection{Uniqueness and the Link to String Theory}
We have established that for any $\phi^3$ Feynman diagram, as well as for any sum of $\phi^3$ Feynman diagrams obtained by piecing together sub-amplitudes, there is a CHY integrand $\mathcal{I}$ such that $\int\!\chyInt\,\mathcal{I}$ is equal to that Feynman diagram or that sum of Feynman diagrams. Furthermore, a simple procedure has been given for finding such particular integrands. The procedure does not, however, work in the reverse. This is because the polygons formed inside a CHY diagram change with the relative distances of the external points, and drawing the external points equidistantly on a circle will not always result in internal polygons that exhibit the Feynman diagrams that the CHY integral evaluates to. We should also stress that the external points of a CHY diagram need to have the correct ordering if the internal polygons in the CHY diagram are to reflect the Feynman diagram. This can be clearly illustrated with the following example:
\eq{\hspace{-20pt}\fig{-56.375pt}{1}{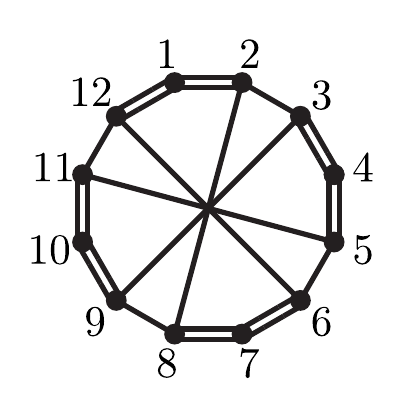}\bbox{=}\fig{-56.375pt}{1}{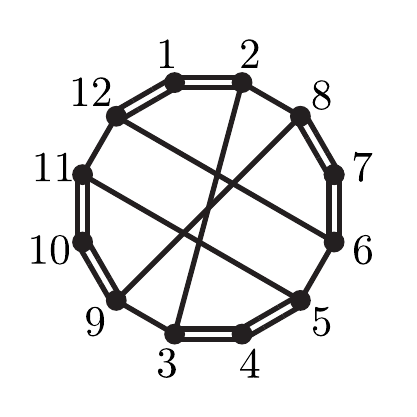}\LR\fig{-56.375pt}{1}{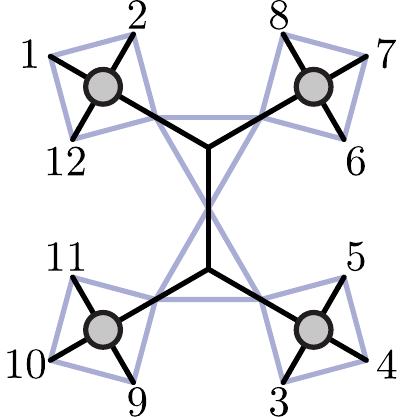}\vspace{-5pt}\label{reodering_fig}}
which evaluates to:
\vspace{10pt}\eq{\displaystyle\frac{\Binv{1,2,12}^{-1}\Binv{3,4,5}^{-1}\Binv{6,7,8}^{-1}\Binv{9,10,11}^{-1}}{\sthree12{12}\sthree345\sthree678\sthree9{10}{11}\ssix12678{12}}\,;\label{reordering}\vspace{5pt}}
here, we have introduced a function $\Binv{i,\ldots,j}$ which corresponds to the {\it inverse} of the $\phi^3$ scattering amplitude involving the legs, 
\vspace{0pt}\eq{\Binv{i,\ldots,j}^{-1}\equiv\mathcal{A}^{\phi^3}\!\!\big(p_i,\ldots,p_j,-(p_i+\cdots+p_j)\big) \,.\label{definition_of_blob_function}}

Strictly speaking, the correspondence between Feynman diagrams and CHY integrands is not one-to-one. Different integrands can in fact yield the same sum of Feynman diagrams upon integration. For example, the diagram,
\eq{\hspace{-40pt}\fig{-56.375pt}{1}{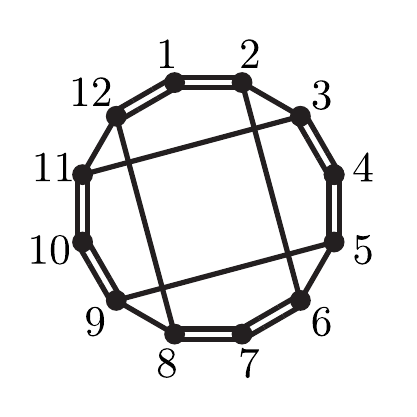}\hspace{-10pt}\bbox{\Rightarrow}\;\displaystyle\frac{\Binv{1,2,12}^{-1}\Binv{3,4,5}^{-1}\Binv{6,7,8}^{-1}\Binv{9,10,11}^{-1}}{\sthree12{12}\sthree345\sthree678\sthree9{10}{11}\ssix12678{12}}\,,\label{nondecomp}}
evaluates to the same expression as (\ref{reodering_fig}). 

However, if we restrict our attention to the CHY diagrams that can be decomposed into two Hamiltonian cycles, then the correspondence does in fact becomes one-to-one. (Notice that the CHY diagram in (\ref{nondecomp}) does not have this property.) All diagrams with this two-cycle property evaluate to the sum of $\phi^3$ Feynman diagrams compatible with the ordering of both the cycles, as was first shown in~\cite{Cachazo:2013iea}.  

In \mbox{ref.\ \cite{Baadsgaard:2015voa}}, we showed that for two-cycle CHY integrands there is a one-to-one correspondence between CHY integrals and string theory integrals in the infinite-tension limit. Without loss of generality, we can assume that one of the two cycles is $(123\cdots n)$, and consider the following CHY integrand:
\begin{align}
\mathcal{I}(z)=\left[\text{Cycle}(z)\prod_{i=1}^N(z_i-z_{i+1})\right]^{-1}\,.
\end{align}
In that case, if one removes the perimeter and replaces the CHY measure with the string theory measure, then in the $\alpha'\!\rightarrow\!0$ limit the string theory integral will exactly evaluate to the CHY integral. Specifically,
\begin{align}
\int\!\chyInt\ \mathcal{I}=\lim_{\alpha'\rightarrow 0} (\alpha')^{n-3} \int d\mu \hspace{1mm} \Lambda(\alpha',p,z) \frac{1}{\text{Cycle}(z)}\,,
\end{align}
where the string theory measure is given by
\begin{align}
d\mu \equiv \hspace{1mm}&\delta(z_A-z_A^0)\delta(z_B-z_B^0)\delta(z_C-z_C^0)\\
&(z_A-z_B)(z_A-z_B)(z_B-z_C)\prod_{i=2}^n\theta(z_{i-1}-z_i)\prod_{i=1}^ndz_i\,,
\end{align}
and where we have defined
\begin{align}
\Lambda(\alpha',p,z) \equiv \prod_{i=1}^{n-1}\prod_{j=i+1}^{n}(z_i-z_j)^{\alpha'\stwo{i}{j}}\,.
\end{align}
So the correspondence between Feynman diagrams and integrands, and the procedure for finding the pertinent integrands, immediately extends to string theory. The subject of expanding integrals out into the contributions arising from individual tree-graphs was in fact explored by Nakanishi already in the early days of the Veneziano model (see {\it e.g}.\ \cite{Nakanishi:1971ve,Nakanishi:1971vf,Nakanishi:1971vi}).

The cycles Cycle$(z)$ that yield a non-zero result upon integration are those that can be represented as a set of polygons meeting at the vertices. The structures of such cycles mirror those of Feynman diagrams of scalar theories---pure triangles corresponding to $\phi^3$, pure squares to $\phi^4$, etc. So we are naturally led to consider how to construct CHY integrands for any kind of scalar field theory.
\\
\\
\\
\section{CHY Representations of $\phi^p$ Scalar Field Theories}\label{sec:phi_p_theories}
Using the knowledge of how CHY diagrams can be represented as Feynman diagrams and the integration rules that we have presented,
we will now show how we can build up $n$-point amplitudes in $\phi^p$ scalar field theories (and mixed versions thereof).

\subsection{The Construction of Scattering Amplitudes in $\phi^p$ Scalar Field Theory}
Let us now describe how to construct CHY representations of scattering amplitudes in $\phi^p$ scalar field theory. Because the combinations of Feynman diagrams contributing to $p$-point sub-amplitudes in $\phi^3$-theory can be represented by polygonal graphs as described above, we may convert these into fundamental $\phi^p$-vertices by including appropriate numerator factors. Representing each of these $\phi^p$ vertices by a $p$-gon, it is clear that the $n$-point scattering amplitude would be represented by all graphs constructed by connecting $p$-gons at their vertices. The full $n$-point amplitude in $\phi^p$-theory would then be obtained by summing over all dihedrally- and reflectionally-distinct ways of gluing $p$-gons together. Examples of the distinct contributions for various amplitudes are summarized in \mbox{Table \ref{polygon_graphs_table}}.

\begin{table}[t]\caption{Distinct polygon graphs contributing to $n$-point amplitudes in $\phi^p$-theories.}\label{polygon_graphs_table}\vspace{-5pt}$$\begin{array}{|@{$\,\,\,$}c@{$\,\,\,$}|@{$\,\,\,$}c@{$\,$}|@{$\,$}c@{$\,$}|@{$\,$}c@{$\,$}|@{}}\multicolumn{1}{@{}c@{}}{n\,\,\,}&\multicolumn{1}{c}{\!\!\!\phi^3\text{-theory}}&\multicolumn{1}{c}{\phi^4\text{-theory}}&\multicolumn{1}{c}{\phi^5\text{-theory}}\\\hline p&\fig{-6.45pt}{1}{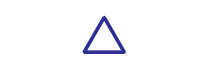}&\fig{-6.45pt}{1}{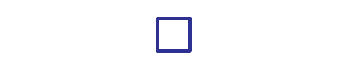}&\fig{-6.45pt}{1}{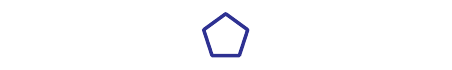}\\\hline 2p-2&\fig{-6.45pt}{1}{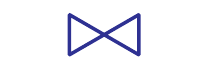}&\fig{-6.45pt}{1}{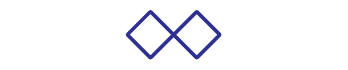}&\fig{-6.45pt}{1}{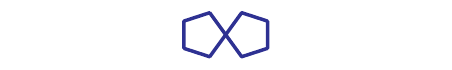}\\\hline 3p-4&\fig{-13.85pt}{1}{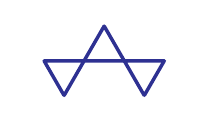}&\fig{-13.85pt}{1}{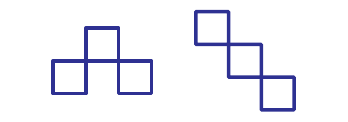}&\fig{-13.85pt}{1}{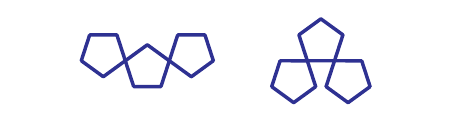}\\\hline 4p-6&\fig{-46.5pt}{1}{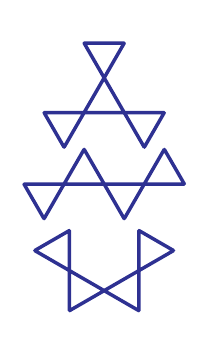}&\fig{-46.5pt}{1}{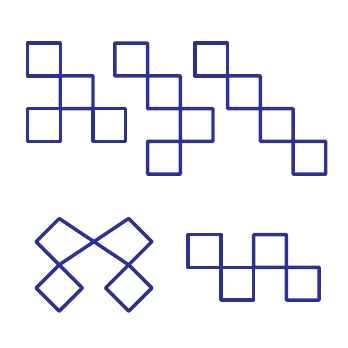}&\fig{-46.5pt}{1}{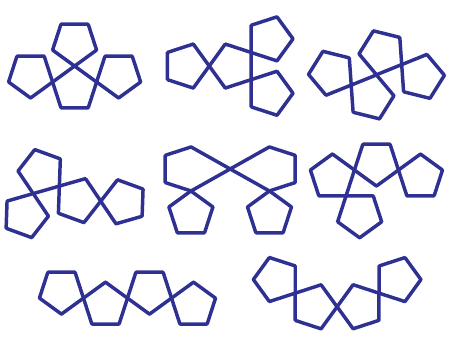}\\\hline\end{array}\vspace{-10pt}$$\end{table}

We can illustrate this general procedure with a few concrete examples. A trivial example would be the $4$-point amplitude in $\phi^4$ theory, which would be generated by a single polygonal graph (a single box):
\eq{\mathcal{A}_{4}^{\phi^4}=\left(\frac{1}{\stwo12}+\frac{1}{\stwo23}\right)^{\!\!\!-1}\!\!\!\fig{-26.5pt}{0.75}{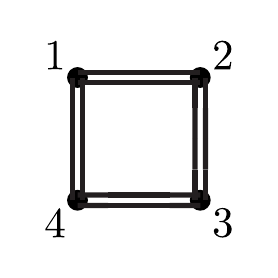}\!\!\!=\,\,\Binv{1,2,3}\!\!\!\fig{-26.5pt}{0.75}{4pt_1}\!\!\!=1\,.}

And fundamental $\phi^4$-vertices can be glued together to form higher-point amplitudes in the obvious way. For the $6$-point amplitude, there is only one (dihedrally-distinct) polygonal graph,
\eq{\phantom{\,.}\fig{-13.85pt}{1}{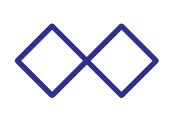}\,.}
This graph corresponds to three distinct contributions to the amplitude:
\eq{\begin{array}{@{}r@{}c@{$\,$}l@{}}\mathcal{A}_{6}^{\phi^4}\!\!=&&\displaystyle\left(\frac{1}{\stwo12}\!+\!\frac{1}{\stwo23}\right)^{\!\!\!-1}\!\!\left(\frac{1}{\stwo45}\!+\!\frac{1}{\stwo56}\right)^{\!\!\!-1}\!\!\!\fig{-26.5pt}{0.75}{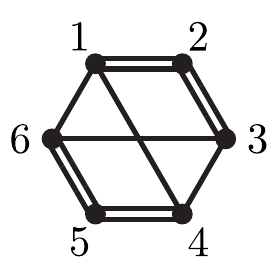}+\text{(2 rotations)},\\=&&\displaystyle\frac{1}{\sthree123}+\frac{1}{\sthree234}+\frac{1}{\sthree345}\,.\end{array}}

Similarly, the $8$-point amplitude in $\phi^5$ theory would be generated by a single polygon diagram, 
\eq{\phantom{\,,}\fig{-13.85pt}{1}{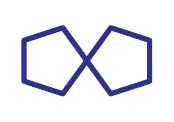}\,,}
corresponding to four dihedrally-distinct contributions to the amplitude,
\eq{\begin{array}{@{}r@{}c@{$\,$}l@{}}\mathcal{A}_{8}^{\phi^5}\!\!=&&\Binv{1,2,3,4}\Binv{5,6,7,8}\!\fig{-26.5pt}{0.75}{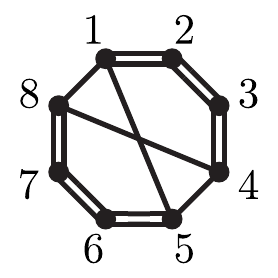}\!+\text{(3 rotations)},\\=&&\displaystyle\frac{1}{\sfour1234}\!+\!\frac{1}{\sfour2345}\!+\!\frac{1}{\sfour3456}\!+\!\frac{1}{\sfour4567}\,.\end{array}}

The generalization to any $n$-point amplitude in $\phi^p$-theory should be quite clear. For example, the $8$-point amplitude in $\phi^4$-theory would be generated by the two polygonal diagrams,
\eq{\phantom{\,.}\fig{-13.85pt}{1}{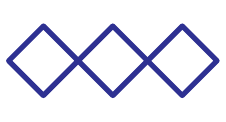},\qquad\fig{-13.85pt}{1}{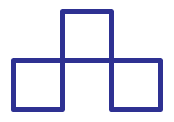}\,.\label{phi4_8pt_weaves}}
The first of these corresponds to the contributions,
\eq{\hspace{-18pt}\begin{array}{@{}r@{}c@{$\,$}l@{}}\mathcal{A}_{8}^{\phi^4\!\!,(1)}\!\!=&&\,\Binv{1,2,3}\Binv{5,6,7}\Binv{8,(1\!+\!2\!+\!3),4}\!\,\fig{-26.5pt}{0.75}{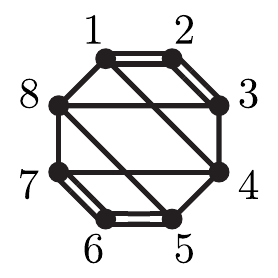}+\text{(3 rotations)},\\=&&\displaystyle\frac{1}{\sthree123\sthree567}\!+\!\frac{1}{\sthree234\sthree678}\!+\!\frac{1}{\sthree345\sthree781}\!+\!\frac{1}{\sthree456\sthree812}\,,\\[12.5pt]\end{array}}
where `$(1\!+\!2\!+\!3)$' denotes the momentum $(p_1\!+\!p_2\!+\!p_3)$; and the second diagram in (\ref{phi4_8pt_weaves}) corresponds to the contributions,
 \eq{\hspace{-18pt}\begin{array}{@{}r@{}c@{$\,$}l@{}r@{}}\mathcal{A}_{8}^{\phi^4\!\!,(2)}\!\!=&&\,\Binv{1,2,3}\Binv{(1\!+\!2\!+\!3),4,5}\Binv{6,7,8}\!\,\fig{-26.5pt}{0.75}{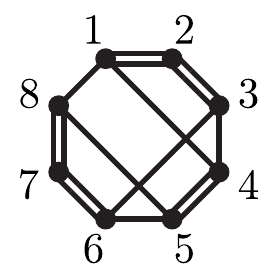}+&\text{(7 rotations)},\\=&&\displaystyle\frac{1}{\sthree123\sthree678}\!+\!\frac{1}{\sthree234\sthree781}\!+\!\frac{1}{\sthree345\sthree812}\!+\!\frac{1}{\sthree456\sthree123}\!+\ldots&(\text{8 terms total}).\\[12.5pt]\end{array}}
Combining these two contributions, we obtain the entire $8$-point amplitude,
\vspace{0pt}\eq{\mathcal{A}_{8}^{\phi^4}=\mathcal{A}_{8}^{\phi^4\!\!,(1)}+\mathcal{A}_{8}^{\phi^4\!\!,(2)}\,.}

For one final example, let us consider the $10$-point amplitude in $\phi^4$-theory. For this amplitude, there are five distinct polygon diagrams that contribute,
\eq{\phantom{}\hspace{-20pt}\fig{-36.375pt}{1}{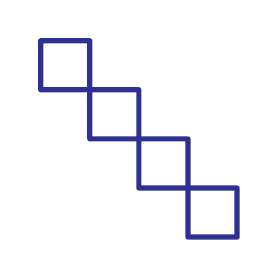}\fig{-36.375pt}{1}{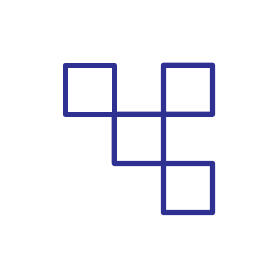}\fig{-36.375pt}{1}{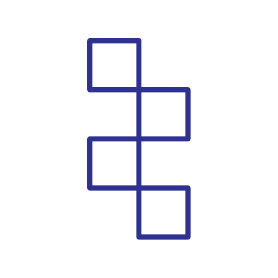}\fig{-36.375pt}{1}{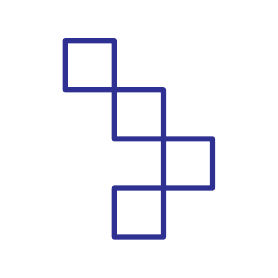}\fig{-36.375pt}{1}{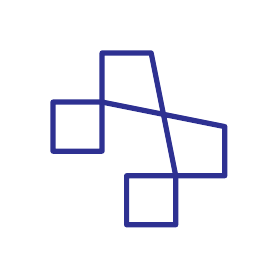}\label{phi4_10pt_weavings}\;}
The first polygon diagram in (\ref{phi4_10pt_weavings}) corresponds to the contributions,
\eq{\hspace{-36pt}\begin{array}{@{}r@{}c@{$\,$}l@{}r@{}}\mathcal{A}_{10}^{\phi^4\!\!,(1)}\!\!=&&\!\!\!\Binv{1,2,3}\Binv{10,(1\!+\!2\!+\!3),4}\Binv{5,(6\!+\!7\!+\!8),9}\Binv{6,7,8}\!\,\fig{-26.5pt}{0.75}{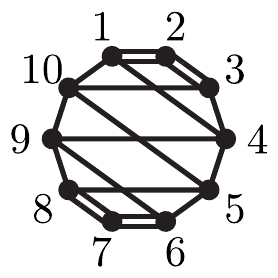}\\&+&(\text{4 rotations})\,=\,\displaystyle\frac{1}{\sthree123\sfive56789\sthree678}+\ldots\,&\hspace{-40pt}(\text{5 terms}).\\[12.5pt]\end{array}}
For the second diagram in (\ref{phi4_10pt_weavings}) we have,
\eq{\hspace{-36pt}\begin{array}{@{}r@{}c@{$\,$}l@{}r@{}}\mathcal{A}_{10}^{\phi^4\!\!,(2)}\!\!=&&\!\!\!\Binv{1,2,3}\Binv{(1\!+\!2\!+\!3),4,(5\!+\!6\!+\!7)}\Binv{5,6,7}\Binv{8,9,10}\!\,\fig{-26.5pt}{0.75}{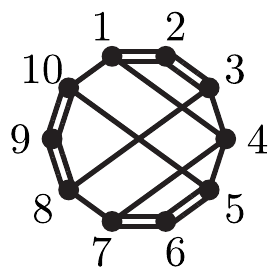}\\&+&(\text{9 rotations})\,=\,\displaystyle\frac{1}{\sthree123\sthree567\sthree89{10}}+\ldots\,&\hspace{-40pt}(\text{10 terms});\\[12.5pt]\end{array}}
for the third, 
\eq{\hspace{-36pt}\begin{array}{@{}r@{}c@{$\,$}l@{}r@{}}\mathcal{A}_{10}^{\phi^4\!\!,(3)}\!\!=&&\!\!\!\Binv{1,2,3}\Binv{(1\!+\!2\!+\!3),4,5}\Binv{6,7,8}\Binv{(6\!+\!7\!+\!8),9,10}\!\,\fig{-26.5pt}{0.75}{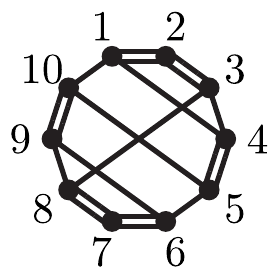}\\&+&(\text{9 rotations and reflections})\,=\,\displaystyle\frac{1}{\sthree123\sfive12345\sthree678}+\ldots\,&\hspace{-40pt}(\text{10 terms});\\[12.5pt]\end{array}}
for the fourth,
\eq{\hspace{-36pt}\begin{array}{@{}r@{}c@{$\,$}l@{}r@{}}\mathcal{A}_{10}^{\phi^4\!\!,(4)}\!\!=&&\!\!\!\Binv{1,2,(3\!+\!4\!+\!5)}\Binv{3,4,5}\Binv{6,(7\!+\!8\!+\!9),10}\Binv{7,8,9}\!\,\fig{-26.5pt}{0.75}{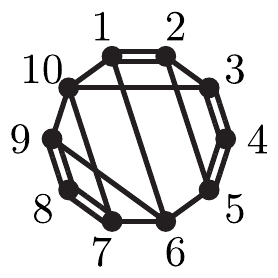}\\&+&(\text{19 rotations and reflections})\,=\,\displaystyle\frac{1}{\sthree345\sfive12345\sthree789}+\ldots\,&\hspace{-40pt}(\text{20 terms});\\[12.5pt]\end{array}}
and for the fifth and final polygon diagram in (\ref{phi4_10pt_weavings}), we have,
\eq{\hspace{-36pt}\begin{array}{@{}r@{}c@{$\,$}l@{}r@{}}\mathcal{A}_{10}^{\phi^4\!\!,(5)}\!\!=&&\!\!\!\Binv{9,10,(1\!+\!2\!+\!3)}\Binv{1,2,3}\Binv{4,5,6}\Binv{(4\!+\!5\!+\!6),7,8}\!\,\fig{-26.5pt}{0.75}{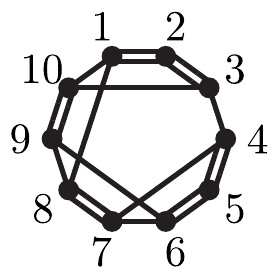}\\&+&(\text{9 rotations})\,=\,\displaystyle\frac{1}{\sthree123\sthree456\sfive45678}+\ldots\,&\hspace{-40pt}(\text{10 terms}).\\[12.5pt]\end{array}}
Thus, the $10$-point amplitude would be represented
\eq{\mathcal{A}_{10}^{\phi^4}=\mathcal{A}_{10}^{\phi^4\!\!,(1)}+\mathcal{A}_{10}^{\phi^4\!\!,(2)}+\mathcal{A}_{10}^{\phi^4\!\!,(3)}+\mathcal{A}_{10}^{\phi^4\!\!,(4)}+\mathcal{A}_{10}^{\phi^4\!\!,(5)}\,.}
\\

\subsection*{Scalar Amplitudes with Mixed Vertices}
From the above considerations it is straightforward to extend the correspondence between Feynman diagrams and CHY integrands to any diagram involving also mixed $m$-vertices $\phi^{p_m}$. The main difference is that the number
of polygon diagrams that has to be considered grows considerably.  This is a consequence of the fact that different polygons can be connected in more ways than identical polygons. 

\clearpage

\subsection{Comparison with the CHY (Pfaffian) Version of $\phi^4$-Theory}
By compactifying the CHY formula for Yang-Mills amplitudes it is possible to obtain the amplitudes of Yang-Mills-scalar theory and as a corollary also a remarkably compact expressions for those of $\phi^4$-theory, \cite{Cachazo:2014xea}. The CHY formula for the (ordered) $\phi^4$ $n$-point function is given by:
\begin{align}
\int\!\chyInt\,\frac{\Pfprime A}{(z_1-z_2)(z_2-z_3)\cdots(z_n-z_1)}\sum_{\substack{\text{connected} \\ \text{perfect matchings $\Xi$}}}\frac{1}{\text{$\Xi$}(z)}\,,
\end{align}
where the matrix $A$ is defined by
\begin{align}
A_{ij}=
\begin{cases}
    \displaystyle\frac{\stwo{i}{j}}{z_i-z_j} & \text{if $i \neq j$}\,,\\
    0 & \text{if $i = j$}\,,
\end{cases}
\end{align}
and where the reduced Pfaffian of $A$ is defined by,
\eq{\Pfprime A\equiv\frac{(-1)^{i+j}}{z_i-z_j}\text{Pf}\,A_{ij}\,,}%
with $A_{ij}$ being the sub-matrix of $A$ obtained by deleting rows and columns $i$ and $j$ of $A$. (This definition of $\Pfprime A$ is in fact independent of the choice of $\{i,j\}$.)
$\Xi(z)$ denotes a ``connected perfect matching''---that is, a product of differences $(z_i-z_j)$ for which every $z_i$ appears in exactly one difference factor and which, if one selects any proper subset $T\!\subset\!\mathbb{Z}_n$ of consecutive numbers, then $\Xi(z)$ will contain at least one factor $(z_k-z_l)$ with $k\!\in\!T$ and $l\!\not\in\!T$.

The CHY $\phi^4$ formula for a general $n$-point function is a far more compact way of writing the amplitude than we can provide with the Feynman diagram 
procedure discussed above. It is also (seemingly) uncorrelated with (and not immediately generalizable to) scalar field theories of arbitrary $\phi^p$-vertices. 

To illustrate the difference, consider the following. If one performs the CHY integration before taking the 
sum over connected perfect matchings $\Xi(z)$, one finds that each term either vanishes or equals one $\phi^4$ Feynman diagram. 
In contrast, in our polygon diagram procedure we evaluate a single CHY integral with the integration rules from \cite{Baadsgaard:2015voa} 
and then effectively cancel out all propagators carrying an even number of external legs with a pre-factor of Mandelstam variables. 
In the CHY formula, the dimensionality is fixed simply by the $\stwo{i}{j}$ factors in $A$, so to obtain a single Feynman diagram 
one must perform the CHY integration over the $(n\mi3)!!$ terms of $\text{Pf}\,A_{ij}$, which non-trivially add up to give the diagram. 
It does present an interesting task to investigate the possible connection between these two formulations further.

\section{Conclusions}\label{sec:conclusions}

We have shown how any scalar amplitude can be represented by a sum 
of polygon diagrams. This construction is very simple and combined with the integration rules for 
CHY from \mbox{ref.\ \cite{Baadsgaard:2015voa}} it offers an alternative to more traditional constructions. Basically, this is a proof of concept on the way to showing that presumably all known field theories can be given a tree-level representation in CHY language. The extension to loop-level remains a challenge, although all the basic ingredients are already in place. Curiously, we are led full circle back to the Feynman diagram expansion, now in an unusual disguise.

Since everything we have done has a very close analogy in string theory it might be interesting to pursue the string theory path of this story in greater detail. An open question is of course
how to construct amplitudes in a similar manner for more complicated cases like Yang-Mills theories. An example could be pure gluon amplitudes where there already
exists a compact CHY construction in terms of a Pfaffian, but where it is an open question if a decomposition 
like the one we have presented for scalar amplitudes here could be worked out as well. Such a formulation 
would likely be closer to a field theory expansion as it would have a direct connection to Feynman diagrams. Issues of gauge invariance need to be faced there.
It is possible that such investigations could open up for further work on the BCJ 
relations~\cite{Bern:2008qj,BjerrumBohr:2009rd, Stieberger:2009hq}. A possible outcome could be a construction
where CHY diagrams corresponding to specific propagator structures could be dressed up 
with numerator factors that satisfy Jacobi identities (a good starting point for such an analysis could be found in {\it e.g}.\ \mbox{refs.\ \cite{BjerrumBohr:2012mg,Monteiro:2013rya}}).

\acknowledgments\label{sec:Acknowledgements} 
This work has been supported in part by a MOBILEX research grant from the Danish
Council for Independent Research (JLB).

\newpage
\providecommand{\href}[2]{#2}\begingroup\raggedright\endgroup

\end{document}